\documentclass[11pt,twoside,onecolumn,a4paper]{article}
\usepackage[]{latexsym}
\usepackage{epsfig}
\usepackage{amsmath,amssymb}
\usepackage[dvips]{color}
\newcommand{\be}{\begin{eqnarray}}
\newcommand{\ee}{\end{eqnarray}}

\newcommand{\lp}{\ell_{\rm P}}
\newcommand{\mpl}{M_{\rm P}}
\renewcommand{\lg}{\ell_{\rm G}}
\newcommand{\mg}{M_{\rm G}}
\newcommand{\rh}{R_{\rm H}}

\title{\bf Brane-world stars and (microscopic) black holes}
\author{
R.~Casadio$^{a,b}$\thanks{casadio@bo.infn.it}
$\ $
and
J.~Ovalle$^{c}$\thanks{jovalle@usb.ve}
\\
\null
\\
$^a${\em Dipartimento di Fisica, Universit\`a di Bologna}
\\
{\em via Irnerio~46, 40126 Bologna, Italy}
\\
\\
$^b${\em Istituto Nazionale di Fisica Nucleare, Sezione di Bologna}
\\
{\em via Irnerio~46, 40126 Bologna, Italy}
\\
\\
$^c${\em Departamento de F\'{\i}sica, Universidad Sim\'on Bol\'{\i}var,}
\\
{\em Apartado 89000, Caracas 1080A, Venezuela}}
\begin{document}
\maketitle
\begin{abstract}
We study stars in the brane-world by employing the principle of minimal geometric
deformation and find that brane-world black hole metrics with a tidal charge
can be consistently recovered in a suitable limit.
This procedure allows us to determine the tidal charge as a function
of the ADM mass of the black hole (and brane tension).
A minimum mass for semiclassical microscopic black holes can then be
derived, with a relevant impact for the description of black hole events at the LHC.  
\end{abstract}
%
%
%
%
%
%
%
%
%
\section{Introduction}
\setcounter{equation}{0}
In the Randall-Sundrum (RS) brane-world (BW) models~\cite{RS}, 
the fundamental scale of gravity $\mg$ can be as low as the
electroweak scale, which would allow for the production of TeV-scale
black holes (BHs) at the Large Hadron Collider (LHC)~\cite{cavaglia}.
One should however not forget that fully analytical metrics describing
BHs in the complete five-dimensional space-time of the RS model~\cite{bwbh}
are yet to be found (albeit, the existence of large static BHs was recently
established in Ref.~\cite{wiseman}).
\par
Solving the full five-dimensional Einstein field equations with appropriate
sources (BH plus brane) is indeed a formidable task (see~\cite{wiseman,cmazza}
and References therein).
The approach we will follow here is therefore to start from a sensible
description of BW stars and employ a limit for (roughly speaking)
vanishing star radius  which (at least formally) allows one to recover the
BH geometry found in Ref.~\cite{dadhich}.
We should warn the reader this method is not free of hindrances.
For example, it was shown in Refs.~\cite{BGM} that the collapse
of a homogeneous star leads to a non-static exterior,
contrary to what happens in four-dimensional General Relativity (GR),
and a possible exterior was later found which is radiative~\cite{dad}
(at the classical level, whereas four-dimensional BHs emit Hawking
radiation~\cite{hawking}, a quantum field theory effect).
If one regards BHs as the natural end-state of the gravitational collapse
of compact objects, one may conclude that classical BHs in the BW
should suffer of the same problem as semiclassical BHs in GR:
no static configuration for their exterior might be allowed~\cite{tanaka,gercas}.
Nonetheless, a static configuration was recently found in Ref.~\cite{wiseman}
(see also Refs.~\cite{dejan,page} for more explicit examples of static metrics),
thus keeping the debate open as to what we should consider a realistic
physical description of BHs in models with extra spatial dimensions.
\par
In any case, one may view a static metric (both in GR and in the BW)
as an approximate description of BHs which holds for sufficiently short times.
This said, although there is no guarantee that our mathematical limit reproduces
the physics of gravitational collapse, we shall find interesting
consequences which may help to better understand BH physics in the BW.
In particular, we shall show that the outer tidal charge $q$~\cite{dadhich}
and ADM mass $\mathcal{M}$ can be both uniquely determined by the BH
proper mass $M$ and brane density $\lambda=\mpl\,\sigma/\lp$.
This, in turn, will allow us to estimate the minimum mass of tidally
charged BHs, thus filling a crucial gap in the arguments of
Ref.~\cite{acmo}.
\par
We shall explicitly display the Newton constant $G_{\rm N}=\lp/\mpl$ and denote
by $\lg\gg\lp$ and $\mg\ll\mpl$ the fundamental five-dimensional length and mass
($\hbar=\mpl\,\lp=\mg\,\lg$).
The ``brane density'' parameter $\sigma$ thus has dimensions of an inverse
squared length, namely $\sigma\simeq\lg^{-2}$.
\section{The brane-world}
\setcounter{equation}{0}
In the RS model, our Universe is a co-dimension one, four-dimensional
hypersurface of vacuum energy density $\lambda$~\cite{RS}.
In Gaussian normal coordinates $x^A=(x^\mu,y)$, where $y$ is the
extra-dimensional coordinate with the brane located at $y=0$
(capitol letters run from $0$~to~$4$ and Greek letters from $0$~to~$3$),
the five-dimensional metric can be expanded near the brane
as~\cite{maart}
\be
g^{(5)}_{AB}
\simeq
\left.g^{(5)}_{AB}\right|_{y=0}
+\left.2\,K_{AB}\right|_{y=0}\, y
+\left.\pounds_{\hat n}K_{AB}\right|_{y=0}\,y^2
\ ,
\label{g5}
\ee
where $K_{AB}$ is the extrinsic curvature of the brane,
and $\pounds_{\hat n}$ the Lie derivative along the unitary four-vector
$\hat n$ orthogonal to the brane.
Junction conditions at the brane lead to~\cite{shiromizu}
\be
K_{\mu\nu}\sim
T_{\mu\nu} -\frac{1}{3}\,\left(T-\lambda\right)\,g_{\mu\nu}
\ ,
\label{K}
\ee
where $T_{\mu\nu}$ is the stress tensor of the matter localized
on the brane, and~\cite{maart}
\be
\pounds_{\hat n} K_{\mu\nu}\sim {\cal E}_{\mu\nu}+F_{\mu\nu}
\ ,
\label{f}
\ee
where ${\cal E}_{\mu\nu}$ is the (traceless) projection of the Weyl tensor on the
brane and $F_{\mu\nu}=F_{\mu\nu}(\lambda,T)$ a tensor which depends on
$T_{\mu\nu}$ and $\lambda$.
\par
We recall that junction conditions in GR allow for surfaces with either step-like
discontinuities or (infinitely thin) Dirac $\delta$-like discontinuities
in the stress tensor~\cite{israel}.
Since a brane in RS is a thin surface, it generates an orthogonal
discontinuity of the extrinsic curvature in five dimensions.
Any discontinuities in the BW matter stress tensor
would induce further discontinuities in the extrinsic curvature~(\ref{K})
parallel to the brane, which should appear in the
five-dimensional metric~(\ref{g5}), and are not allowed by the regularity of
five-dimensional geodesics.
Moreover, because of the term of order $y^2$ in Eq.~(\ref{g5}),
and Eq.~(\ref{f}), the projected Weyl tensor must also be continuous
on the brane.
This regularity requirements can be understood by considering that,
in a microscopic description of the BW, matter should be smooth
along the fifth dimension, yet localized on the brane (say, within a width
of order $\sigma^{-1/2}\simeq\lg$, see Ref.~\cite{g2}).
In order to build a physically acceptable model of BW stars,
continuity of five-dimensional geodesics should then be preserved
by smoothing the matter stress tensor and the projected Weyl
tensor across the star surface~\cite{gercas}.
Since this renders the problem nearly intractable, we shall apply
a simplifying idea which has already proved very effective.
\par
%
%
%
%
%
%
We shall start from the effective four-dimensional equations of
Ref.~\cite{shiromizu}.
In general, these equations form an open system on the brane,
because of the contribution ${\cal E}_{\mu\nu}$ from the bulk Weyl tensor
in Eq.~\eqref{f},
and identifying specific solutions requires more information on
the bulk geometry and a better understanding
of how our four-dimensional space-time is embedded in the bulk.
Nonetheless, it is possible to generate the BW version of every 
spherically symmetric GR solution through the
{\it minimal geometric deformation approach\/}~\cite{jovalle2009}. 
This method is based on requiring that BW solutions reduce
to four-dimensional GR solutions at low energies ({\em i.e.}, energy
densities much smaller than $\sigma$).
From the point of view of a brane observer,
the five-dimensional gravity therefore induces a geometric deformation
in the radial metric component, which is the source of anisotropy
on the brane.
When a solution of the four-dimensional Einstein equations
is considered as a possible solution in the BW, 
the geometric deformation produced by extra-dimensional effects
is minimal, and the open system of effective BW equations is automatically
satisfied.
This approach was successfully used to generate physically
acceptable interior solutions for stellar systems~\cite{jovalle07},
and even exact solutions were found in Ref.~\cite{jovalle207}.
\subsection{Brane-world star exterior}
The metric outside a spherically symmetric BW source should
solve the effective four-dim\-ensional vacuum Einstein
equations~\cite{shiromizu},
\be
R_{\mu\nu}-\frac{1}{2}\,R^\alpha_{\ \alpha}\,g_{\mu\nu}
=
\mathcal{E}_{\mu\nu}
\qquad
\Rightarrow
\qquad
R^\alpha_{\ \alpha}=0
\ ,
\ee
where we recall that tidal effects are represented by $\mathcal{E}_{\mu\nu}$.
Only a few analytical solutions are known~\cite{wiseman,dejan,page,fabbri}, 
one being the tidally charged metric~\cite{dadhich} 
\be
ds^2
=
e^{\nu}\,dt^2-e^{\lambda}\,dr^2+r^2\left(d\theta^2+\sin^2\theta\,d\phi^2\right)
\ ,
\label{metric}
\ee
with $\lambda\equiv\lambda_+=-\nu\equiv-\nu_+$, and
\be
e^{\nu_+}=1-\frac{2\,\lp\,{\cal M}}{\mpl\,r}-\frac{q}{r^2}
\ ,
\label{tidalg}
\ee
which has been extensively studied in Refs.~\cite{acmo,bcLHC}.
The tidal charge $q$ and ADM mass ${\cal M}$ are usually treated as
independent quantities.
However, by studying the interior BW solution of compact stars, 
it will be possible to obtain a relationship between $q$ and ${\cal M}$
from the usual junction conditions~\cite{israel}. 
For instance, the simplest case of a static, spherically symmetric and
uniform distribution with $\mathcal{E}_{\mu\nu}=0$ has already been
considered in Ref.~\cite{MaartensGermani2001}, where the tidal charge $q$
and ADM mass ${\cal M}$ were then shown to be related, that is
$q=q(M,R,\rho,\sigma)$ and ${\cal M}={\cal M}(M,\rho,\sigma)$,
with $M$ the total GR mass of a star of radius $R$ and density $\rho$.
There are however an additional free constant parameter $\alpha$ and a 
constraint $f(M,R,\rho,\sigma,\alpha)=0$ relating these functions
at the star surface in a non trivial way, thus making it impossible to display
a simple relationship between the tidal charge $q$ and ADM mass ${\cal M}$.
In our case, a more realistic BW distribution, with $\mathcal{E}_{\mu\nu}\neq\,0$,
is considered in such a way that these difficulties are overcome,
and simple expressions for the tidal charge and ADM mass will
be obtained (see Eqs.~\eqref{Madm} and \eqref{qM} below).
\par
First of all, if no source is present and ${\cal{M}}=0$, or in the GR limit
$\sigma^{-1}\to 0$, the tidal charge $q$ should vanish.
Let us therefore assume $q=q({\cal M},\sigma)$.
We then consider the junction conditions at the star surface
$r=R$ between the metric given in Eq.~({\ref{metric}) and
a general interior solution of the form~\eqref{metric}
with $\nu=\nu_-(r)$ and $\lambda=\lambda_-(r)$ to be found
through the minimal geometric deformation approach.
In this approach, from the point of view of a brane observer,
the five-dimensional gravity produces a geometric deformation
in the radial metric component of a general spherically symmetric
metric~\eqref{metric} given by
\be
e^{-\lambda}
=
1-\frac{2\,\lp\,m(r)}{\mpl\,r}
+
\underbrace{\rm Geometric\ Deformation}_{=f(\nu,\rho,p)}
\ ,
\label{expect}
\ee
where $m(r)$ is the standard GR mass function,
\be
m(r)
=
4\pi\,\int_0^r \rho\,x^2\,dx
\ ,
\label{GRm} 
\ee
with $\rho$ the matter energy density.
When a solution of the four-dimensional Einstein field equations
with density $\rho$ and pressure $p$ is considered as a possible BW solution,
the geometric deformation in Eq.~\eqref{expect} is minimal, 
and given by
\be
f^{*}(r)
=
\frac{8\,\pi}{\sigma}\,
e^{-I}\,\int_0^r
\frac{2\,x\,e^I}{x\,\nu'+4}\left(\rho^2+3\,\rho\, p\right)
dx
\ ,
\label{fsolutionmin}
\ee
with
\be
I
\equiv
\int\frac{2\,r\,\nu''+r\,{\nu'}^2+{4\,\nu'}+\frac{4}{r}}
{r\,\nu'+4}\,dr
\ .
\label{I}
\ee
Hence, any perfect fluid solution is altered by the geometric deformation $f^{*}(r)$
produced by five-dimensional effects at high energies.
This geometric deformation produces imperfect fluid effects 
through the BW solution for the geometric function $\lambda(r)$, 
which can be written as
\be
e^{-\lambda}=1-\frac{2\,\tilde{m}(r)}{r}
\ ,
\label{reglambda}
\ee
where the interior mass function $\tilde{m}$ is given by
\be
\tilde{m}(r)
=
m(r)-\frac{r}{2}\,f^{*}(r)
\ .
\label{massfunction}
\ee
\par
Using a generic interior solution $\nu_-=\nu(r)$ and the $\lambda_-=\lambda(r)$ shown
in Eq.~\eqref{reglambda}, along with the tidally charged metric~\eqref{tidalg},
in the matching conditions at the stellar surface $r=R$, we find
\be
\label{mgd00}
&&
e^{\nu_R}
=1-\frac{2\,\lp\,\cal{M}}{\mpl\,R}-\frac{q}{R^2}
\ ,
\\
\label{RegmatchNR1}
&&
\frac{2\,\cal{M}}{R}
=
\frac{2\,M}{R}-\frac{\mpl}{\lp}\left(f^{*}+\frac{q}{R^2}\right)
\ ,
\\
\label{RegmatchNR2}
&&
\frac{q}{R^4}
=\left(\frac{{\nu}'_R}{R}+\frac{1}{R^2}\right)f^{*}+8\,{\pi}\,\frac{\lp}{\mpl}\,p_R
\ ,
\ee
where $\nu_R\equiv \nu_-(R)$, $\nu'_R\equiv \partial_r\nu_-|_{r=R}$,
$M=m(R)$ is the total GR mass of the star,
$p_R$ the pressure at the surface and $f^{*}=f^*(R;M,\sigma)$ encodes
the minimal geometric deformation undergone by 
the radial metric component due to bulk effects~\cite{jovalle2009}
at $r=R$.
In particular, the tidal charge $q$ can be obtained from
the second fundamental form of the surface $r=R$,
which in the BW is given by
\be
p_R
+\frac{1}{\sigma}
\left(\frac{\rho_R^2}{2}+\rho_R\,p_R
+\frac{2}{k^4}\,{\cal U}_R^-\right)
+\frac{{4\,\cal P}_R^-}{k^4\,\sigma}
=
\frac{2\,{\cal U}_R^+}{k^4\,\sigma}
+\frac{4\,{\cal P}_R^+}{k^4\,\sigma}
\ ,
\label{sffxx}
\ee
where $\pm$ denote the limits $r\to R^\pm$ respectively.
In our approach, the above reduces to
\be
\frac{\lp}{\mpl}\,p_R
+\left(\frac{\nu'_R}{R}
+\frac{1}{R^2}\right)
\frac{f^*}{8\,\pi}
= 
\frac{2\,{\cal U}_R^+}{k^4\,\sigma}
+\frac{4\,{\cal P}_R^+}{k^4\,\sigma}
\ .
\label{sffmgd}
\ee
On using the tidally charged metric~\eqref{tidalg} and 
\be
{\cal U}^+_R
=
-\frac{{\cal P}^+_R}{2}
=
\frac{4\,\pi\, q\,\sigma}{3\,R^4}
\ee
in Eq.~\eqref{sffmgd}, we find the expression shown in
Eq.~\eqref{RegmatchNR2}.
From Eqs.~(\ref{RegmatchNR1}) and~(\ref{RegmatchNR2}),
we then obtain the tidal charge as
\be
\frac{\mpl}{\lp}\,q
=
\left(\frac{R\,\nu'_R+1}{R\,\nu'_R+2}\right)
\left(\frac{2\,M}{R}-\frac{2\,{\cal M}}{R}\right)
R^2
+\frac{8\,\pi\,p_R\,R^4}{2+R\,\nu'_R}
\ .
\label{qgeneral}
\ee
This charge $q$ only depends on the interior structure through
$\nu'_R$.
We may therefore choose a suitable $\nu'_R$ in order to obtain
$q=q({\cal M},\sigma)$.
\par
For instance, by taking $p_R=0$ and imposing the boundary 
constraint (at $r=R$) 
\be
 R\,\nu'_R
=
-\frac{(M-{\cal M})-\frac{2\,{\cal M}\,K\,\mpl}{\sigma\,R^2\,\lp}}
{(M-{\cal M})-\frac{{\cal M}\,K\,\mpl}{\sigma\,R^2\,\lp}}
\ ,
\label{bound}
\ee
where $K$ is a (dimensionful) constant we can fix later,
we obtain a simple relation between $q$ and ${\cal M}$ given by
\be
q
=
\frac{2\,K\,{\cal M}}{\sigma\,R}
\ ,
\label{qpeculiar}
\ee
where we must have 
\be
\frac{{\cal M}\,K\,\mpl}{\sigma\,R^2\,\lp}
<
(M-{\cal M})
<
\frac{2\,{\cal M}\,K\,\mpl}{\sigma\,R^2\,\lp}
\label{cond}
\ee
to ensure an acceptable physical behaviour in the interior
[see Eq.~(\ref{bound})].
Remarkably, this expression for $q$ satisfies all of our requirements,
namely:
\begin{description}
\item[a)]
it vanishes for ${\cal M}\to 0$ and for $\sigma^{-1}\to 0$, and
\item[b)]
it vanishes for very small star density, that is for $R\to\infty$ at fixed
$\cal M$ and $\sigma$.
\end{description}
Let us further notice that $q$ diverges for $R\to 0$, at fixed $\cal M$ and $\sigma$,
which one may expect from previous considerations about point-like singularities in
the BW.
The condition~(\ref{bound}) then leads to the simple exterior solution
\be
e^{\nu}
=
1
-\frac{2\,\lp\,{\cal M}}{\mpl\,r}
\left(1+\frac{\mpl\,K}{\lp\,\sigma\,R\,r}\right)
\ .
\label{simple}
\ee
\par
It is interesting to note that the solution~(\ref{simple})
can also be obtained without imposing any {\em ad hoc\/}
boundary constraints of the form in Eq.~(\ref{bound}). 
We can in fact employ the result that the pressure does not
need to vanish at the surface in the BW.
From Eq.~(\ref{qgeneral}), we can then find a value for $p_R$
leading to Eq.~(\ref{simple}), namely
\be
{4\,\pi\,R^3\,p_R}
=
\frac{\mpl\,{\cal M}\,K}{\lp\,\sigma\,R^2}
\left(2+R\,\nu'_R\right)
-(M-{\cal M})
\left(1+R\,\nu'_R\right)
\ .
\label{bound2}
\ee
Note that $p_R\to 0$ for $R\to\infty$ (at fixed $M$, $\mathcal M$
and $\sigma$), as well as for $\sigma^{-1}\to 0$ (at fixed $R$),
if ${\mathcal M}\to M$ in the same limit [in fact, see Eq.~(\ref{Madm}) below].
\subsection{Brane-world star interior}
The star radius $R$ is still a free parameter.
However, it can be fixed for any specific star interiors.
For instance, let us consider the exact interior BW solution found in
Ref.~\cite{jovalle207},
\be
e^{\nu}=A\left(1+C\,r^2\right)^4
\ ,
\ee
where $A$ and $C$ are constants.
Using this metric element, with BW matter density
\be
\rho
=
C_\rho
\left(\frac{\mpl}{\lp}\right)
\frac{C\left(9+2\,C\,r^2+C^2\,r^4\right)}
{7\,\pi \,{\left( 1 + C\,r^2 \right) }^3}
\ ,
\ee
where $C_\rho=C_\rho(K)$ is also a constant to be determined for consistency,
and vanishing surface pressure
\be
p_R
=
\left(\frac{\mpl}{\lp}\right)
\frac{2\,C\left(2-7\,C\,R^2-C^2\,R^4\right)}{7\,\pi\left(1+C\,R^2\right)^3}
=0
\ ,
\ee
leads to $C=\frac{\sqrt{57}-7}{2\,R^2}$, and, from Eq.~\eqref{GRm},
we find
\be
R=
2\,n
\left(\frac{\lp}{\mpl}\right)
\frac{M}{C_\rho}
\ ,
\label{exac4}
\ee
with $n \equiv \frac{56}{43-\sqrt{57}}\simeq 1.6$.
This result, with~\footnote{Different choices of $K$ (and corresponding $C_\rho$)
are also being considered and will be reported elsewhere.}
\be
K
=
\left(\frac{\mpl}{\mg}\right)^2
\frac{\lg}{\mg}
\ ,
\ee
can be used in the constraint~(\ref{bound}) to obtain
$C_\rho=\left({\mg}/{\mpl}\right)^4$ and the ADM mass
\be
{\cal M}
=
\frac{M^3}
{M^2+n_1\,\mg^2}
\ .
\label{Madm}
\ee
Moreover, the expression for the tidal charge~\eqref{qpeculiar}
yields
\be
q
=
\frac{\lg^2\,M^2}
{n\,\left(M^2+n_1\,\mg^2\right)}
\ ,
\label{qM}
\ee
where we used $\sigma\simeq\lg^{-2}$ and
$
{n}_1
\equiv
\frac{1}{2\,n^2}
\left(\frac{5\,C\,R^2+1}{9\,C\,R^2+1}\right)
=
\frac{31653-1007\,\sqrt{57}}{175616}
\simeq
0.14
$.
Upon inverting the relation~\eqref{Madm} between ${\mathcal M}$ and $M$, 
one could finally express the tidal charge $q=q({\mathcal M},\sigma)$.
The explicit expression of $M=M({\mathcal M})$ is however rather cumbersome 
and we shall not display it here.
Moreover, the expressions~(\ref{exac4}) and~(\ref{Madm}) satisfy the
bounds in Eq.~\eqref{cond}, and are therefore associated with
a consistent boundary condition~(\ref{bound}).
\section{Black hole limit and minimum mass}
\setcounter{equation}{0}
The ADM mass~(\ref{Madm}) and tidal charge~(\ref{qM}) do not explicitly
depend on the star radius $R$, and we can therefore assume they are
valid in the limit $R\to 0$ (or, more cautiously, $R\ll\lg$).
Correspondingly, the exterior metric given by Eq.~\eqref{simple} becomes
\be
e^{\nu}
=
1
-\frac{2\,\lp\,M^3}
{\mpl\left(M^2+n_1\,\mg^2\right) r}
\left(1
+\frac{\lg^2\,\mpl}
{2\,n\,\lp\,M\,r}
\right)
\ ,
\label{simple2}
\ee 
which can be used to describe a BH of ``bare'' (or proper) mass $M$.
\begin{figure}[t]
\center
\includegraphics[scale=0.8]{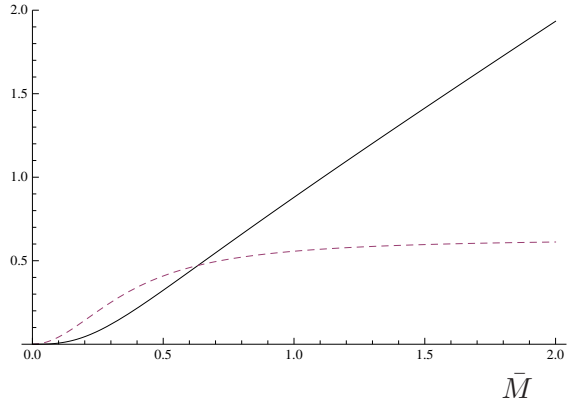}
\\
\hspace{6cm}$\bar M$
\centering\caption{Dimensionless ADM mass $\bar{\cal M}$ from Eq.~(\ref{bMM})
(solid line) and charge $\bar q$ from Eq.~(\ref{bq}) (dashed line) {\em vs\/} the
``bare'' mass $\bar M$.}
\label{fig1}      
\end{figure}
\begin{figure}[h]
\center
\raisebox{4cm}{$\bar q$}
\includegraphics[scale=0.8]{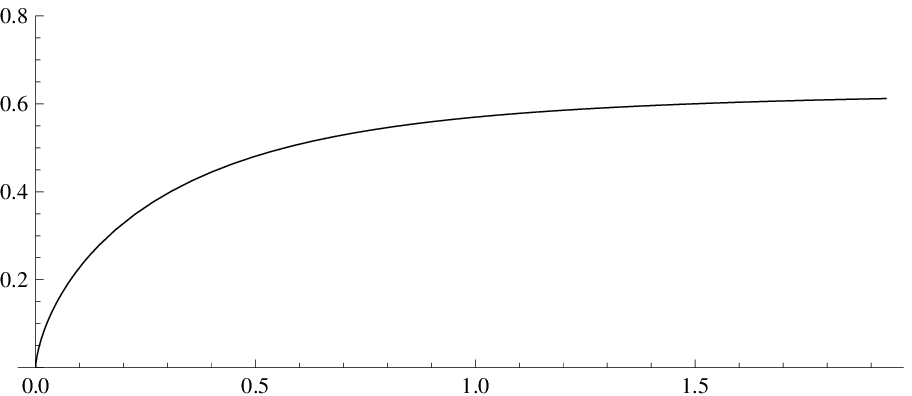}
\\
\hspace{6cm}$\bar{\mathcal M}$
\centering\caption{Dimensionless charge $\bar q$ vs ADM mass $\bar{\cal M}$. }
\label{fig2}       
\end{figure}
Fig.~\ref{fig1} shows the dimensionless ADM mass
\be
\bar{\cal M}
=
\frac{\cal M}{\mg}
=
\frac{\bar M^3}
{\bar M^2+{n}_1}
\simeq
\frac{\bar M^3}
{0.1+\bar M^2}
\ ,
\label{bMM}
\ee
and tidal charge
\be
\bar q
=
\frac{q}{\lg^2}
=
\frac{\bar M^2}
{n\,({n}_1+\bar M^2)}
\simeq
\frac{\bar M^2}{0.2+1.6\,\bar M^2}
\ ,
\label{bq}
\ee
as functions of the dimensionless proper mass
$\bar M={M}/{\mg}$.
Fig.~\ref{fig2} shows the dimensionless charge $\bar q$ as a function of the ADM mass
$\bar{\cal M}$. 
Note that, for $M\gtrsim\mg$, the ADM mass ${\cal M}\simeq M$, whereas
the tidal charge saturates to a maximum
\be
q_{\rm max}\simeq 0.6\,\lg^2
\ ,
\ee
and is practically negligible for macroscopic BHs.
\par
From Eq.~\eqref{qM}, we can now infer an important result for microscopic BHs.
The condition that is usually taken to define a semiclassical BH is its horizon
radius $\rh$ be larger than the Compton wavelength
$\lambda_M\simeq \lp\,\mpl/M$ (see~\cite{acmo} and References therein).
From Eq.~(\ref{tidalg}), we obtain the horizon radius
\be
\rh
=
\frac{\lp}{\mpl}\left({\cal M}+\sqrt{{\cal M}^2+q\,\frac{\mpl^2}{\lp^2}}\right)
\ ,
\ee
and the classicality condition $\rh\gtrsim\lambda_M$ reads
\be
\frac{M}{\mpl^2}
\left({\cal M}+\sqrt{{\cal M}^2+q\,\frac{\mpl^2}{\lp^2}}\right)
\gtrsim
1
\ .
\ee
We define the critical mass $M_c$ as the value which saturates the above bound.
In order to proceed, we shall expand for $M\sim{\cal M}\simeq \mg\ll\mpl$,
thus obtaining
\be
\frac{\rh^2}{\lambda_{M}^2}
\simeq
\frac{M^2}{\mpl^2}\frac{q}{\lp^2}
\simeq
\frac{\mg^2}{\mpl^2}\,\bar M^2\,\bar q\,\frac{\lg^2}{\lp^2}
\simeq
\bar M^2\,\bar q
\simeq
1
\ ,
\ee
or $\bar M^4\simeq {n\left({n}_1+\bar M^2\right)}$,
which yields
\be
M_c\simeq 1.3\,\mg
\ ,
\ee
or ${\mathcal M}_c\simeq 1.2\,\mg$, from Eq.~\eqref{bMM}.
This can be viewed as the minimum allowed mass for a semiclassical
BH in the BW.
\section{Conclusions and outlook}
\setcounter{equation}{0}
We have analyzed analytical descriptions of stars in the BW
to recover the BH metric~(\ref{tidalg}) of Ref.~\cite{dadhich}, with the tidal charge
as an explicit function of the ADM mass and brane tension,
which was still an open problem.
Using the most general junction conditions between a generic interior
solution and the tidally charged metric~(\ref{tidalg}) in the minimal geometric
deformation approach~\cite{jovalle2009}, a simple relationship among
the tidal charge, the ADM mass and the brane tension satisfying all physical
requirements was found.
The simple solution~(\ref{qM}) for the tidal charge $q=q({\cal M},\sigma)$
then allowed us to determine the minimum mass of a semiclassical microscopic
BH, namely $\mathcal{M}_c\simeq 1.2\,\mg$.
\par
Our results could be relevant for the description of BH events at the LHC.
For example, the fact that $\mathcal{M}_c\gtrsim\mg$ means that producing BHs,
even within TeV-scale gravity, might be beyond the LHC capability (for some
data analysis of BH events at the LHC, see Refs.~\cite{CMS}).
%
%
%
%
%
%
%
%
%
%
%
%

%
\end{document}